# Evolving scientific collaboration among EU member states, candidate countries and global partners: 2000-2024

**Corresponding author:**
Myroslava Hladchenko, Center for R&D Monitoring , Faculty of Social Sciences,  University of Antwerp, Antwerp, Belgium
hladchenkom@gmail.com
Myroslava.Hladchenko@uantwerpen.be

**Purpose**
This study explores how EU integration, globalisation, and geopolitical disruptions have influenced scientific collaboration among European countries at different stages of EU membership. Specifically, it distinguishes between the EU-14, the EU-13, that joined the EU in 2004 or later, and EU candidate countries.
**Design/methodology**
Using Scopus article, the study analyses Relative Intensity of Collaboration (RIC) among EU member state, candidate countries and China, Latin America, the UK, the USA and Russia.
**Findings**
Findings indicate increasing integration within European groups and with global partners, yet persistent hierarchical structures remain. EU-14 countries form the core of the network, exhibiting stable and cohesive collaboration, including with the UK despite Brexit. EU-13 countries occupy an intermediate position, showing moderate collaboration with EU-14 but stronger collaboration within their own group, with EU candidate countries and Russia. EU candidate countries demonstrate even weaker integration with EU-14, focusing on intra-group ties and links with EU-13 and Russia. RIC peaks in 2012 and 2018 for EU-13 and EU candidate countries correspond to Horizon 2020 and Horizon Europe cycles, highlighting the role of EU Framework Programmes. Collaboration with Russia increased following 2014 and only marginally declined after 2022. For EU-14, it exceeds collaboration with the USA. Collaboration with China remains limited due to network and cultural constraints, with similar intensity across all three groups. Overall, funding and policy initiatives are critical for stable international collaboration.
**Research limitations**
The analysis is limited by the Scopus database coverage.
**Policy implications**
Findings suggest that to strengthen the EU's scientific position, policymakers should prioritise targeted funding and strategic initiatives that bridge collaboration gaps.
**Originality/value**
This study provides a comprehensive, longitudinal analysis of European scientific collaboration, highlighting hierarchical structures, the differential roles of EU-14, EU-13, and candidate countries, and the resilience of networks with global partners such as the UK and Russia, while linking collaboration dynamics to EU Framework Programmes.

**1 Introduction**
Scientific collaboration is widely recognised as a fundamental driver of knowledge creation and technological innovation (Adams, 2013; Wuchty et al., 2007). Studying international co-authorship provides insights into how research systems are embedded in broader socio-political structures and global science dynamics (Melin, 2000; Beaver, 2001; Persson et al., 2004). Understanding the intensity and structure of cross-border collaboration is crucial for policymakers, funding agencies, and university managers aiming to strengthen research capacity and integration (Makkonen & Mitze, 2016).

In Europe, co-authorship patterns have long been shaped by geopolitical and institutional contexts, and shifts in political regimes. Border changes, and integration processes have reshaped landscape of scientific collaboration (Glänzel & Schubert, 2001). Foremost, this refers to the fall of the socialist bloc (Igić, 2002; Kozak et al., 2015; Glänzel et al., 1999) and the establishment and enlargement of the European Union (Kwiek, 2021; Makkonen & Mitze, 2016; Teodorescu and Andrei, 2011). Europe's enlargements—through the accession of new member states and the establishment of Association Agreements with candidate countries—were intended not only to promote political and economic alignment but also to integrate research systems (Luukkonen et al., 1992). EU Framework Programmes such as FP7 (2007-2013), Horizon 2020 (2014-2020), and Horizon Europe (2021-2027) have played a central role in encouraging global scientific collaboration among EU member states, promoting engagement with third countries through joint calls, international partnerships, and co-funded research initiatives. (European Commission, 2015; Veugelers, 2021). Science has become more globalised than ever before, with collaboration increasingly spanning continents and cultures (Wagner & Leydesdorff, 2005; Ganzi et al., 2012; Glänzel, 2001; Kwiek, 2021). European researchers are embedded in a dense web of global partnerships, with collaboration intensity often correlating with research impact (Glänzel & Schubert, 2004; Abramo et al., 2009). However, recently, Brexit and Russo-Ukrainian war have emerged as geopolitical disruptions with potential implications for European scientific collaboration (Makkonen & Mitze, 2023).

This study explores how EU integration, globalisation, and geopolitical disruptions have influenced scientific collaboration among European countries at different stages of EU membership. Specifically, it distinguishes between the EU-14: long-standing Western and Southern European member states prior to the 2004 enlargement (Austria, Belgium, Denmark, Finland, France, Germany, Greece, Ireland, Italy, Luxembourg, Netherlands, Portugal, Spain, Sweden); the EU-13: the Central and Eastern European countries that joined the EU in 2004 or later (Austria, Belgium, Denmark, Finland, France, Germany, Greece, Ireland, Italy, Luxembourg, Netherlands, Portugal, Spain, Sweden); and EU candidate countries (Albania, Bosnia and Herzegovina, Georgia, Moldova, North Macedonia, Serbia, Turkey, Ukraine). These groups reflect differing historical trajectories, institutional capacities, and levels of integration into European and global research networks.

The study contributes in three ways. First, it enhances understanding of the complex dynamics of European research collaboration by examining the Relative Intensity of Collaboration (RIC) across three groups of countries: EU-14, EU-13, and EU candidate countries. By comparing collaboration trends over a 25-year period, the study assesses whether EU enlargement and related policy instruments have effectively fostered balanced research integration. Second, it examines the intensity of the EU's global research connections with major scientific partners, including China, the United States, and Latin America. Third, it investigates how major disruptions, such as Brexit and the Russo-Ukrainian war, have affected collaboration with the UK and Russia. Overall, the findings aim to inform ongoing debates on European research integration, the effectiveness of EU science policy, and the resilience of scientific collaboration in times of geopolitical uncertainty.

**2 EU Framework Programmes as a drive of within EU integration**
The main assumption underlying EU research and innovation policy is that an integrated European innovation system, characterised by effective network structures and system-wide knowledge diffusion, is key to sustainable technological and economic competitiveness (European Commission, 2007). Founding EU countries such as France, Germany, Belgium, and the Netherlands have a long-lasting collaboration due to geographical, historical, and cultural proximity (Glänzel & Schubert, 2004; Hoekman et al., 2010). In addition, their scientific integration was stimulated by EU early financial initiatives, specifically by the establishment of the European Framework Programme in 1984, by the European Communities (EC) and involved the ten European Economic Community member states.

FP7 (2007–2013) strengthened scientific collaboration between EU founding countries and Central and Eastern European countries, such as Poland, the Czech Republic, Hungary, and the Baltic states, that joined EU in 2004 and 2007 (Scherngell & Lata, 2013; Makkonen & Mitze, 2016**;** Teodorescu & Andrei, 2011; Glänzel & Schlemmer, 2007). The fostered scientific collaboration was supposed to deepen integration among old and new EU states and strengthen intra-European ties, making the Union a cohesive and competitive research community in the global landscape (Hoekman et al., 2009; Ulnicane, 2015). FP7 was followed by Horizon 2020 and Horizon Europe. Horizon 2020 introduced specific measures to strengthen the integration of Widening countries into the Framework Programme, targeting primarily most EU-13 and Associated Countries lagging behind (European Court of Auditors, 2022). In Horizon2020, the Widening countries also included Luxembourg and Portugal, while in Horizon Europe they included Greece and Portugal from EU-15. It resulted in strengthened collaboration between EU-15 and EU-13 countries (EC, 2025). To further strengthen collaboration with Widening countries, the European Commission established the WIDERA (Widening Participation and Strengthening the European Research Area) component within Horizon Europe (EC, 2025). WIDERA calls typically require project consortia to be coordinated by institutions from Widening countries. The Horizon Europe budget for widening actions tripled compared to Horizon 2020.

Despite some progress in the success rate of Widening countries in obtaining funding through Horizon Europe, during the first two years of the programme, Germany alone proportionally received more than all the Widening countries. It implies that participation in EU programmes should be complemented by national governments reforming the R&I sector (Kelk & Drake, 2023). Efficient knowledge exchange and enhancement of connectivity of the Widening countries occur through collaboration with EU 'bridge' countries that facilitate these processes. There are dense clusters in co-authorship networks linking older and newer EU and associated members, often anchored by Germany, France, Italy, Spain and the UK as key collaboration hubs (European Commission, 2025). Germany has always been the most important cooperation partner for East European scientific communities (Glänzel & Schubert, 2005).

**EU member states and global research collaboration partners**
European Framework Programmes encourage collaboration not only within EU member states and EU candidate countries but also with global partners. EU member states participate automatically in FP, while other countries must sign a Horizon Association Agreement in order to take part under equivalent conditions. Low- and middle-income countries may participate as partners and can receive EU funding, whereas high-income non-associated countries typically participate on a self-funded basis. Norway and Switzerland are associated with Horizon Europe. Despite Brexit, with a transition period spanning 2016–2020, the UK remained associated with Horizon 2020 and was later granted association status to Horizon Europe. By contrast, the USA, is not associated with FP, though it is at the core of the global scientific network (Olechnicka et al., 2019; Gui et al., 2019; Marini, & Mouritzen, 2025) and it has been the biggest international collaborative partner for China, the UK, Germany, France, Italy, Spain and the Netherlands (Kwiek, 2019). Such high-income countries as Australia, Brazil, Canada, Chile, China, Japan, South Korea are also not associated with Horizon Europe.

Scientific collaboration between the European Union (EU) and Latin America has expanded steadily over the past two decades, driven by shared priorities in research and innovation (Belli & Nenoff, 2022). Spain and Portugal have historically been the strongest partners of Latin America due to linguistic ties and historical connections (Chinchilla-Rodríguez et al., 2016; Russell et al., 2020; Lemarchand, 2012). Overall, scientific ties between the EU and Latin America reflect both regional priorities and the growing need to address global challenges through collaborative research (Ronda-Pupo, 2024).

Since the first decade of the 21st century, alongside the USA, China has gained a prominent position as the central hub of the global scientific collaboration network (Baker, 2023; Wu et al., 2024; Hu et al., 2020). EC has implented joined calls with Japan, Brazil and China have implemented jointly funded calls with China in FP7, Horizon 2020 and Horizon Europe (European Commission, Directorate-General for Research and Innovation, n.d.).Overall, since the 2000s, EU-China scientific co-authorship has grown rapidly, driven by major joint projects in fields such as renewable energy, engineering, and materials science. Germany, the UK, and the Netherlands are among China's top European collaborators (Gómez-Espés et al, 2024; Wang et al., 2017). However, at the end of 2025, the European Commission announced the exclusions of Chinese universities, including the so-called "Seven Sons of National Defence," from participating in roughly half of Horizon Europe, particularly in clusters related to health, digital technologies, and civil security, citing security and dual-use concerns (Matthews, 2025). India and South Africa participate as international partners and are eligible for EU funding. In 2022, the European Commission suspended Russian public bodies from participation in Horizon 2020 and Horizon Europe projects, terminating their involvement in ongoing grants and halting payments to these entities (EC, 2022)

Overall, the exploration of knowledge networks of EU member states and EU candidate countries revealed that they are inclined to cooperate with countries that are in close geographical or have similar strategic interests and socio-economic conditions (European Commission, 2025). This aligns with a global trend of geographical, socioeconomic, political, linguistic and cultural proximities fostering international collaboration (Hoekman et al., 2009; Glänzel & Schubert, 2004; European Commission, 2025). Consequently, it results in the establishment of regional clusters (Chessa et al., 2013). For example, Balkan candidate countries (Serbia, Montenegro, North Macedonia, Albania) often have strong bilateral ties with EU members like Croatia, Slovenia, or Hungary due to historical, linguistic, or cultural proximity and cross-border projects (European Commission 2025).

## 3 Data and methods
### 3.1 Relative Intensity of Collaboration (RIC)

Relative Intensity of Collaboration (RIC), a widely used metric for quantifying the strength of research collaborations (Coccia & Bozeman, 2016; Fuchs et al., 2021; Luukkonen et al., 1992). This study uses RIC as the primary measure to assess the intensity of bilateral and multilateral international collaborations, allowing us to systematically compare EU state members collaborative activity across different partner countries and groups.

The Relative Intensity of Collaboration (RIC) (Fuchs et al., 2021) was used as a main metric in this study. The RIC indicator compares the share of country Y within the collaboration profile of country X to the share of Y within all collaborations in the whole network. The Relative Intensity of Collaboration (RIC) is calculated as the ratio of the observed share of co-authored publications between two countries to the expected share based on their total publication outputs (Fig 1). The RIC indicator is asymmetric and increases in value if collaboration between two countries increases. In this study, the publication-based interpretation of the RIC indicator developed by Fuchs et al., (2021) was used. RIC values of 1.0 indicates that the collaboration between two countries occurs as frequently as expected, based on their total publication outputs and the overall network structure, essentially reflecting random mixing. A value greater than 1.0 suggests preferential or stronger-than-expected collaboration, while a value below 1 indicates weaker-than-expected ties. This approach aligns with established practices in bibliometric studies of international co-authorship (Katz & Martin, 1997).

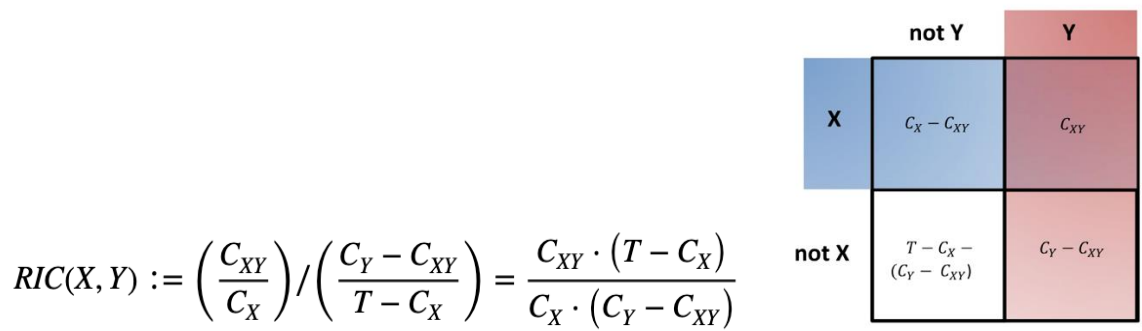

$$RIC(X,Y) := \left(\frac{C_{XY}}{C_X}\right) / \left(\frac{C_Y - C_{XY}}{T - C_X}\right) = \frac{C_{XY} \cdot (T - C_X)}{C_X \cdot (C_Y - C_{XY})}$$

**Figure 1.** Diagram and formular of Relative Intensity of Collaboration (RIC) (Fuchs et a., 2021).

### 3.2 Data
This study examines the Relative Intensity of Collaboration (RIC) between 2000 and 2024 among EU member states and EU candidate countries, both within and across these groups, as well as with selected global partners, namely the United Kingdom, the United States, China, Russia, and Latin America.

Data for this study were drawn from the CWTS in-house version of the Scopus database. First, articles published in Scopus-indexed journals between 2000 and 2024 were identified in the CWTS Scopus database. Second, affiliation countries were assigned to each article. Based on the RIC formula described above, the corresponding code was implemented in the CWTS Scopus database. The RIC indicator was calculated annually for each of EU members states and EU candidate countries with respect to all collaboration partners (EU-14, EU-13, EU candidate countries, the UK, the USA, China, Russia, and Latin America) over the period 2000–2024. EU-14, EU-13, EU candidate countries, and Latin America were treated as aggregated collaboration partners. When a focal country belonged to the respective collaboration group, its articles were excluded from the group total to avoid to avoid self-collaboration. For example, when calculating the RIC between Austria and the EU-14 group, Austria was excluded from the EU-14 collaboration group. Articles were counted using a full-counting approach and were not fractionalised across countries.

Latin American countries included Mexico, Guatemala, Honduras, Nicaragua, El Salvador, Costa Rica, Panama, Belize, Cuba, Dominican Republic, Argentina, Bolivia, Brazil, Chile, Colombia, Ecuador, Paraguay, Peru, Uruguay and Venezuela.

### 3.3 Statistical tests

To analyse long-term patterns in the Relative Intensity of Collaboration (RIC), we estimated a linear mixed-effects regression model (Model 1) capturing temporal dynamics and cross-national heterogeneity in scientific collaboration. The dataset comprises repeated yearly observations (2000–2024) for 35 EU member and candidate countries across multiple collaboration partner groups, forming an unbalanced longitudinal panel. Montenegro was excluded from the dataset because extremely high RIC values were driven by a very small number of internationally co-authored publications, which could bias the analysis.

Given the hierarchical structure of the data, with yearly observations clusterwithin countries, random intercepts were included to control for unobserved, time-invariant country-specific characteristics. The model includes fixed effects for publication year (centred), country group (EU-14, EU-13, EU candidate countries), and collaboration partner (EU-14, EU-13, EU candidate countries, China, Russia, the UK, the USA, and Latin America). To capture heterogeneous temporal dynamics, the specification incorporates a three-way interaction between publication year, country group, and collaboration partner.

Country heterogeneity was modelled through random intercepts, allowing baseline collaboration intensity to vary across countries while avoiding the large number of parameters required by country fixed effects. This specification reduces multicollinearity and facilitates generalisation beyond the sampled countries. Model parameters were estimated using restricted maximum likelihood (REML). The analysis was conducted in R using the *lmer* function from the *lme4* package. The model specification was: RIC ~ pub_year_c * collaboration_partner * group + (1 | country). Model fit was evaluated using marginal and conditional R² (R2m and R2c), which quantify the variance explained by fixed effects alone and by the combined fixed and random effects, respectively.

To evaluate the impact of the Russo–Ukrainian war on collaboration with Russia, we fitted a linear mixed-effects model incorporating interactions between publication year, collaboration group, collaboration partner, and war phase, with random intercepts for countries (Model 2). The war was operationalised as a categorical variable distinguishing the pre-war period (up to 2013), the period of hybrid war following the 2014 annexation of Crimea (2014–2021), and the full-scale Russo-Ukrainian war period (2022 onward). This specification allows the estimation of both gradual temporal trends and structural shifts in collaboration intensity associated with different stages of the conflict.To assess changes in collaboration patterns across key historical milestones, we compared model-predicted RIC values using estimated marginal means for the final pre-war year (2013), the last year prior to the full-scale Russo-Ukrainian war (2021), and the later war period (2024). The model was specified in R as: RIC ~ pub_year_c × collaboration_group × group + war_phase × pub_year_c × collab_group × group + (1 | country).

# 4 Results
## 4.1 Descriptive statistics: mean RIC

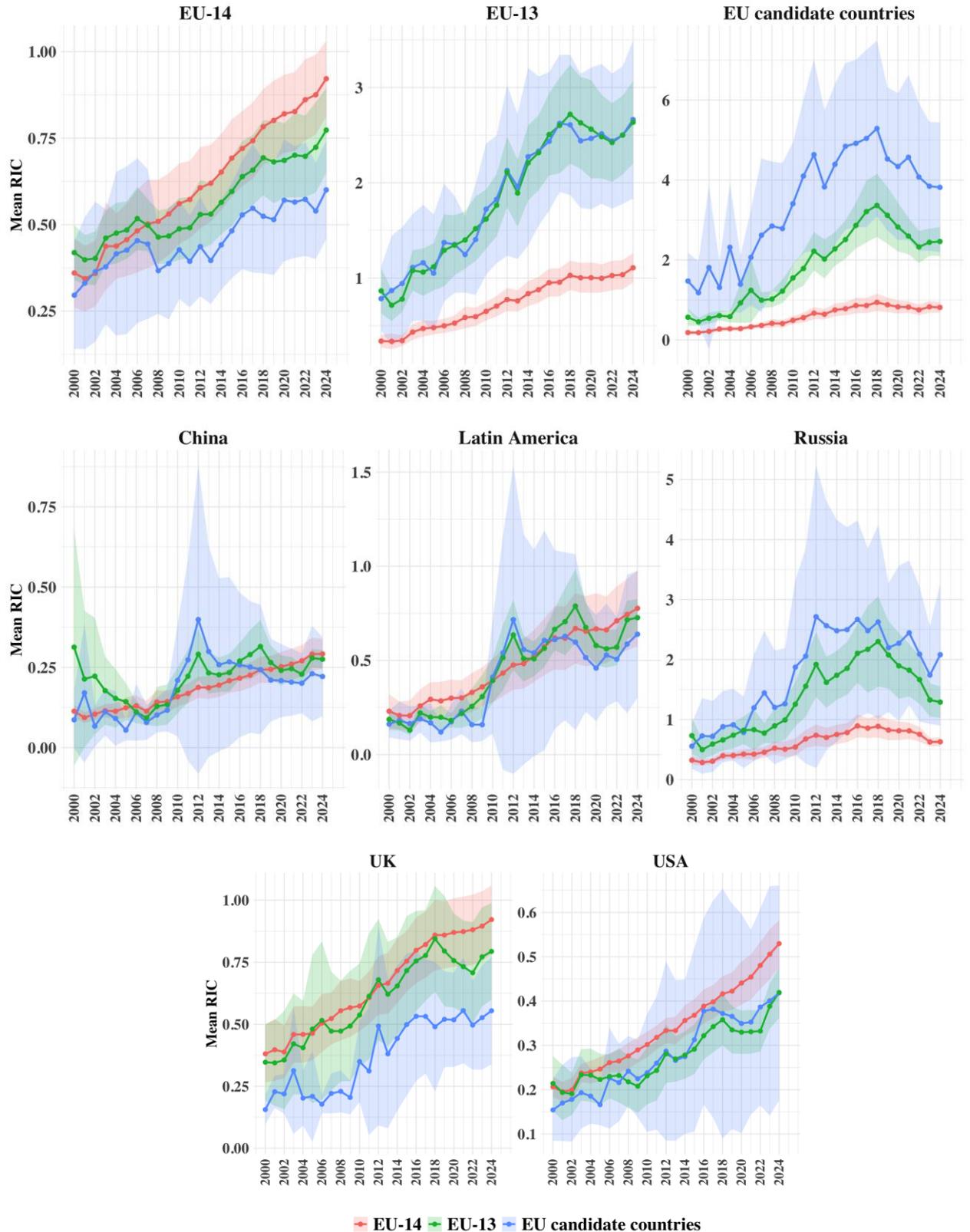

**Figure 2.** Mean RIC with 95% CI by collaboration partner, 2000–2024

Figure 2 illustrates the evolution of mean RIC across EU groups and with major global partners between 2000 and 2024, revealing distinct integration trajectories within EU and global research collaboration networks.

EU-14 countries demonstrate the most stable and structurally consolidated collaboration patterns, achieving the highest mean RIC values in 2024 within the group (0.92), with EU-13 (1.11), and with the United Kingdom (0.92). Although EU-14 countries also show gradual growth in collaboration with the USA (0.53) and Latin America (0.78), these relationships remain comparatively moderate in intensity, with substantially weaker collaboration observed with China (0.29). Collaboration with Russia declines from 0.81 in 2021 to 0.63 in 2024, suggesting partial contraction following full-scale Russo-Ukrainian war, although collaboration intensity remains higher than with the USA.

Collaboration between EU-13 and EU-14 countries demonstrates greater volatility and, after 2007, falls below intra-EU-14 collaboration levels. EU-13 countries collaborate more intensively within their own group and with EU candidate countries than with EU-14, pointing to the formation of a partially integrated collaboration cluster that remains less connected to the main EU research core. Similar dynamics are observed for EU candidate countries, whose collaboration patterns display the highest volatility, reflecting them experiencing stronger instability in collaboration.

Temporal peaks in collaboration intensity are observed around 2012 and 2018 for EU-13 and EU candidate countries, both within and between these groups. These peaks coincide with the late phases of FP7 (2007–2013) and Horizon 2020 (2014-2020), highlighting that EU funding cycles play a critical role in shaping collaboration intensity. Comparable temporal peaks are also visible in collaboration with China, Latin America, the UK, and the USA, indicating that EU framework programmes facilitate global collaboration expansion.

EU-13 and EU candidate countries maintain slightly lower collaboration intensity with the USA compared to EU-14, but levels comparable to EU-14's collaboration with China. Conversely, collaboration with Russia is substantially stronger for EU-13 and particularly for EU candidate countries, suggesting the persistence of regional, linguistic and socio-economic linkages. Although collaboration with Russia declines for EU-13 after 2023, intensity levels remain above the expected baseline, while EU candidate countries show a partial rebound in 2024. Russo-Ukrainian war started in 2014, and it became full-scale in 2022. Notably, from 2014 to 2021 mean RIC between EU-14 and Russia increased from 0.75 to 0.82, between EU-13 and Russia it increased from 1.74 to 1.82 and between EU candidate countries and Russia it rose from 2.48 to 2.08.

Finally, variation in confidence intervals highlights structural differences in collaboration stability. EU-14 countries exhibit relatively narrow mean RIC intervals, indicating consistent collaboration patterns, whereas EU-13 and especially EU candidate countries display wider intervals, reflecting greater heterogeneity across countries.

Overall, the results indicate that EU research collaboration is structured in several layers, including a stable group of highly connected countries, a group of countries that are increasingly integrating into the system, and a group of emerging partners whose collaboration patterns are more variable.

## 4.2 Statistical tests

The results of mixed-effects Model 1 are presented in Table 1. The model demonstrates strong explanatory power for variation in RIC, with the marginal $R^2$ indicating that fixed effects—publication year, country group, collaboration partner, and their interactions—explain approximately 58.5% of the variance in RIC. When accounting for country-level heterogeneity via random intercepts, the conditional $R^2$ increases to 65.8%, underscoring the importance of persistent national differences in shaping collaboration patterns.

Using EU-14 as the reference category, the intercept (0.332, $p < 0.001$) represents the expected RIC for EU-14 countries collaborating within the EU-14 baseline group at the centered publication year. The positive and statistically significant coefficient for publication year (0.024, $p < 0.001$) indicates that RIC increases over time for EU-14 collaborations, reflecting a general upward trend.

Compared with intra-EU-14 collaborations, collaborations with EU candidate countries (-0.150, p = 0.089) or Latin America (-0.159, p = 0.072) show marginally lower impact, while collaboration with China (-0.246, p = 0.006) is associated with much lower RIC.

Three-way interactions indicate that EU candidate countries collaborating within the group (0.131, p < 0.001), with EU-13 (0.066, p < 0.001) or with Russia (0.07, p < 0.001) experience faster growth in collaboration intensity than the EU-14 baseline. The same is observed about EU-13 collaborating within the group (0,064, p < 0.001), with EU candidate countries (0,092, p < 0.001 ) and with Russia (0,047, p < 0.001)

**Table 1** Fixed-effect estimates from a linear mixed-effects model for RIC (2000–2024) by publication year, collaboration partner, group, and their interactions, with random intercepts for country.

| Predictor | Estimate | Std. Error | DF | Pr(>|t|) | Significant |
|---|---|---|---|---|---|
| (Intercept) | 0.332 | 0.097 | 88.55274 | 0.001 | *** |
| Year | 0.024 | 0.004 | 6834.01 | <0.001 | *** |
| China | -0.246 | 0.089 | 6834.156 | 0.006 | ** |
| EU candidate countries | -0.15 | 0.088 | 6834.044 | 0.089 | . |
| EU-13 | -0.012 | 0.088 | 6834.01 | 0.891 | |
| Latin America | -0.159 | 0.088 | 6834.071 | 0.072 | . |
| Russia | 0.017 | 0.089 | 6834.074 | 0.85 | |
| UK | 0.029 | 0.088 | 6834.01 | 0.741 | |
| USA | -0.153 | 0.088 | 6834.01 | 0.083 | . |
| EU candidate countries (group) | 0.019 | 0.16 | 88.55274 | 0.907 | |
| EU-13 (group) | 0.054 | 0.139 | 88.55274 | 0.702 | |
| Year: China | -0.016 | 0.006 | 6834.098 | 0.013 | * |
| Year: EU candidate countries | 0.009 | 0.006 | 6834.035 | 0.149 | |
| Year: EU-13 | 0.011 | 0.006 | 6834.01 | 0.094 | . |
| Year: Latin America | 0.001 | 0.006 | 6834.049 | 0.873 | |
| Year: Russia | -0.002 | 0.006 | 6834.052 | 0.793 | |
| Year: UK | 0.001 | 0.006 | 6834.01 | 0.896 | |
| Year: USA | -0.011 | 0.006 | 6834.01 | 0.09 | . |
| Year: EU candidate countries (group) | -0.015 | 0.007 | 6834.01 | 0.042 | * |
| Year: EU-13 (group) | -0.009 | 0.006 | 6834.01 | 0.148 | |
| China: EU candidate countries | 0.089 | 0.155 | 6834.896 | 0.565 | |
| EU candidate countries: EU candidate countries | 1.43 | 0.147 | 6834.057 | <0.001 | *** |
| EU-13: EU candidate countries | 0.49 | 0.146 | 6834.018 | 0.001 | *** |
| Latin America: EU candidate countries | -0.003 | 0.151 | 6834.466 | 0.983 | |
| Russia: EU candidate countries | 0.498 | 0.149 | 6834.229 | 0.001 | *** |
| UK: EU candidate countries | -0.217 | 0.147 | 6834.045 | 0.138 | |
| USA: EU candidate countries | -0.052 | 0.146 | 6834.01 | 0.722 | |
| China: EU-13 | 0.045 | 0.128 | 6834.167 | 0.726 | |
| EU candidate countries: EU-13 | 0.207 | 0.127 | 6834.026 | 0.103 | |
| EU-13: EU-13 | 0.407 | 0.127 | 6834.01 | 0.001 | *** |
| Latin America:EU-13 | -0.096 | 0.128 | 6834.105 | 0.452 | |
| Russia:EU-13 | 0.241 | 0.128 | 6834.134 | 0.06 | . |
| UK: EU-13 | -0.06 | 0.127 | 6834.01 | 0.634 | |
| USA:EU-13 | -0.053 | 0.127 | 6834.01 | 0.677 | |
| Year: China: EU candidate countries | 0.009 | 0.011 | 6834.53 | 0.415 | |
| Year: EU candidate countries: EU candidate countries | 0.131 | 0,01 | 6834.043 | <0.001 | *** |
| Year: collab_groupEU-13: EU candidate countries | 0.066 | 0,01 | 6834.015 | <0.001 | *** |
| Year: Latin America: EU candidate countries | 0.01 | 0.011 | 6834.277 | 0.356 | |
| Year: Russia: EU candidate countries | 0.07 | 0.011 | 6834.131 | <0.001 | *** |
| Year: UK: EU candidate countries | 0.008 | 0.01 | 6834.034 | 0.446 | |
| Year: USA: EU candidate countries | 0.013 | 0.01 | 6834.01 | 0.219 | |
| Year: China: EU-13 | 0.004 | 0.009 | 6834.106 | 0.63 | |
| Year: EU candidate countries: EU-13 | 0.092 | 0.009 | 6834.022 | <0.001 | *** |
| Year: EU-13:EU-13 | 0.064 | 0.009 | 6834.01 | <0.001 | *** |
| Year: Latin America: EU-13 | 0.011 | 0.009 | 6834.073 | 0.23 | |
| Year: Russia: EU-13 | 0.047 | 0.009 | 6834.091 | <0.001 | *** |
| Year: UK: EU-13 | 0.005 | 0.009 | 6834.01 | 0.566 | |
| Year: USA:EU-13 | 0.004 | 0.009 | 6834.01 | 0.653 | |

Significance levels:***p< 0.001,**p < 0.01, *p < 0.05

Table 2 contains predicted RIC values (estimated marginal means, EMMEANS) for 2000 and 2024, illustrating how collaboration intensity varies by partner group over time. From the EU perspective, intra-EU-14 collaborations (EU-14 × EU-14) show moderate predicted RIC growth, increasing from 0.33 in 2000 to 0.91 in 2024. Collaborations between EU-14 and EU-13 countries increased significantly from 0.32 to 1.15, indicating strong integration. In contrast, collaborations between EU-13 and EU-14 rose more modestly from 0.39 to 0.74, while intra-EU-13 collaborations increased markedly from 0.79 to 2.91. Collaboration between EU-14 and EU candidate countries increased from 0.18 to 0.97, whereas collaboration between EU candidate countries and EU-14 rose from 0.35 to 0.57. Overall, EU-14 have consolidated collaboration both within the group and with EU-13 and candidate countries. Conversely, EU-13 countries demonstrate stronger growth in collaboration within their own group and with candidate countries than in integration with EU-14. A similar pattern is observed for EU candidate countries. These results suggest differentiated integration processes within the EU research network, where newer member states and candidate countries increasingly collaborate among themselves rather than fully converging toward EU-14 collaboration patterns. Regarding the UK, EU-14 collaboration intensity exceeds intra-EU-14 level, while collaboration between EU-13 and the UK remains comparatively lower.

From a global perspective, although EU-14 strengthened collaboration with all non-EU partners, the achieved intensity is comparable to intra-EU-14 level only with Russia (0.88), moderate with Latin America (0.77) and the USA (0.50), and relatively weak with China (0.28). EU-13 and EU candidate countries display lower collaboration intensity with the USA (0.37 and 0.41, respectively), but substantially higher intensity with Russia (2.10 and 2.71, respectively), and comparable levels of collaboration with China (0.27 and 0.25) and Latin America (0.77 and 0.66).

**Table 2** Predicted RIC by collaboration partner in 2000 and 2024 (estimated marginal means with 95% CI)

| Group | Collaboration partner | RIC 2000 | CI lower 2000 | CI upper 2000 | RIC 2024 | CI lower 2024 | CI upper_2024 |
|---|---|---|---|---|---|---|---|
| EU-14 | EU-14 | 0.33 | 0.14 | 0.52 | **0.91** | 0.72 | 1.10 |
|  | China | 0.09 | -0.11 | 0.28 | 0.28 | 0.09 | 0.47 |
|  | EU candidate countries | 0.18 | -0.01 | 0.37 | 0.97 | 0.78 | 1.16 |
|  | EU-13 | 0.32 | 0.13 | 0.51 | 1.15 | 0.96 | 1.34 |
|  | Latin America | 0.17 | -0.02 | 0.36 | 0.77 | 0.58 | 0.96 |
|  | Russia | 0.35 | 0.16 | 0.54 | 0.88 | 0.69 | 1.07 |
|  | UK | 0.36 | 0.17 | 0.55 | 0.96 | 0.77 | 1.14 |
|  | USA | 0.18 | -0.01 | 0.37 | 0.50 | 0.31 | 0.69 |
| EU-13 | EU-14 | 0.39 | 0.19 | 0.58 | 0.74 | 0.54 | 0.93 |
|  | China | 0.18 | -0.02 | 0.38 | 0.27 | 0.07 | 0.46 |
|  | EU candidate countries | 0.44 | 0.24 | 0.64 | 3.22 | 3.02 | 3.41 |
|  | EU-13 | 0.78 | 0.58 | 0.98 | 2.91 | 2.72 | 3.11 |
|  | Latin America | 0.13 | -0.07 | 0.33 | 0.77 | 0.57 | 0.97 |
|  | Russia | 0.64 | 0.44 | 0.84 | 2.10 | 1.90 | 2.29 |
|  | UK | 0.35 | 0.16 | 0.55 | 0.85 | 0.65 | 1.05 |
|  | USA | 0.18 | -0.02 | 0.38 | 0.37 | 0.18 | 0.57 |
| EU candidate countries | EU-14 | 0.35 | 0.10 | 0.60 | 0.57 | 0.32 | 0.82 |
|  | China | 0.19 | -0.08 | 0.46 | 0.25 | -0.01 | 0.50 |
|  | EU candidate countries | 1.63 | 1.38 | 1.88 | 5.21 | 4.96 | 5.46 |
|  | EU-13 | 0.83 | 0.58 | 1.08 | 2.87 | 2.62 | 3.12 |
|  | Latin America | 0.19 | -0.07 | 0.45 | 0.66 | 0.41 | 0.92 |
|  | Russia | 0.86 | 0.61 | 1.12 | 2.71 | 2.46 | 2.97 |
|  | UK | 0.16 | -0.09 | 0.41 | 0.59 | 0.34 | 0.84 |
|  | USA | 0.15 | -0.10 | 0.40 | 0.41 | 0.16 | 0.66 |

Table 3 presents Model 2 results, showing predicted RIC values (estimated marginal means) by country group and collaboration partner across war phases. The model explains substantial variation in RIC: fixed effects account for 60.8% of the variance (marginal $R^2$), increasing to 68.1% when country-level random intercepts are included, highlighting the role of persistent national differences in collaboration patterns.

The model-predicted RIC values indicate that, between 2013 and 2021, collaboration with Russia increased for EU-14 countries from 0.70 to 0.85 and for EU-13 countries from 1.59 to 2.03, while it remained stable at 2.35 for EU candidate countries. This upward trend was followed by a decline between 2021 and 2024: for EU-14 countries, collaboration intensity decreased from 0.85 to 0.63, although it remained higher than EU-14 collaboration with the USA (0.53). For EU-13 countries, collaboration with Russia declined from 2.03 to 1.29, but remained higher than collaboration with EU-14 countries (0.77). For EU candidate countries, collaboration intensity decreased from 2.35 to 2.08.

**Table 3** Predicted RIC by collaboration partner in 2013, 2021 and 2024 (estimated marginal means with 95% CI)

| Group | Collaboration partner | 2013 | CIL2013 | CIU2013 | 2021 | CIL2021 | CIU2021 | 2024 | CIL2024 | CIU2024 |
|---|---|---|---|---|---|---|---|---|---|---|
| EU-14 | EU-14 | 0.62 | 0.41 | 0.83 | 0.84 | 0.60 | 1.09 | 0.92 | 0.59 | 1.26 |
| | China | 0.18 | -0.03 | 0.39 | 0.26 | 0.02 | 0.51 | 0.29 | -0.05 | 0.63 |
| | EU c.c. | 0.62 | 0.41 | 0.83 | 0.88 | 0.63 | 1.12 | 0.81 | 0.48 | 1.15 |
| | EU-13 | 0.77 | 0.55 | 0.98 | 1.04 | 0.80 | 1.28 | 1.11 | 0.77 | 145 |
| | Latin America | 0.46 | 0.25 | 0.68 | 0.69 | 0.45 | 0.93 | 0.78 | 0.44 | 1.11 |
| | **Russia** | **0.70** | 0.48 | 0.91 | **0.85** | 0.60 | 1.09 | **0.63** | 0.29 | 0.97 |
| | UK | 0.66 | 0.45 | 0.87 | 0.90 | 0.65 | 1.14 | 0.92 | 0.59 | 1.26 |
| | USA | 0.34 | 0.12 | 0.55 | 0.45 | 0,21 | 0.70 | 0.53 | 0.19 | 0.87 |
| EU-13 | EU-14 | 0.53 | 0.31 | 0.75 | 0.72 | 0.46 | 0.97 | 0.77 | 0.42 | 1.12 |
| | China | 0.18 | -0.04 | 0.40 | 0.27 | 0.01 | 0.52 | 0.28 | -0.07 | 0.62 |
| | EU c.c. | 1.97 | 1.75 | 2.19 | 3.04 | 2.79 | 3.30 | 2.46 | 2.11 | 2.81 |
| | EU-13 | 1.96 | 1.74 | 2.18 | 2.66 | 2.40 | 2.91 | 2.63 | 2.29 | 2.98 |
| | Latin America | 0.51 | 0.29 | 0.73 | 0.65 | 0.40 | 0.91 | 0.73 | 0.38 | 1.08 |
| | **Russia** | **1.59** | 1.37 | 1.81 | **2.03** | 1.78 | 2.28 | **1.29** | 0.94 | 1.64 |
| | UK | 0.64 | 0.42 | 0.86 | 0.79 | 0.54 | 1.05 | 0.79 | 0.44 | 1.14 |
| | USA | 0.26 | 0.04 | 0.48 | 0.35 | 0.10 | 0.60 | 0.42 | 0.07 | 0.77 |
| EU candidate countries | EU-14 | 0.44 | 0.16 | 0.72 | 0.56 | 0.24 | 0.88 | 0.59 | 0.14 | 1.03 |
| | China | 0.27 | -0.03 | 0.56 | 0.20 | -0.12 | 0.53 | 0.22 | -0.22 | 0.67 |
| | EU c.c. | 4.14 | 3.86 | 4.42 | 4.65 | 4.32 | 4.97 | 3.82 | 3.37 | 4.26 |
| | EU-13 | 1.97 | 1.69 | 2.25 | 2.56 | 2.24 | 2.88 | 2.66 | 2.22 | 3.11 |
| | Latin America | 0.52 | 0.23 | 0.80 | 0.51 | 0.19 | 0.84 | 0.64 | 0.19 | 1.08 |
| | **Russia** | **2.35** | 2.07 | 2.64 | **2.35** | 2.03 | 2.67 | **2.08** | 1.64 | 2.53 |
| | UK | 0.36 | 0.08 | 0.64 | 0.54 | 0.22 | 0.87 | 0.55 | 0.11 | 1.00 |
| | USA | 0.27 | -0.01 | 0.55 | 0.38 | 0.05 | 0.70 | 0.42 | -0.03 | 0.86 |

## 4.3 The two highest RIC peaks, 2000-2024

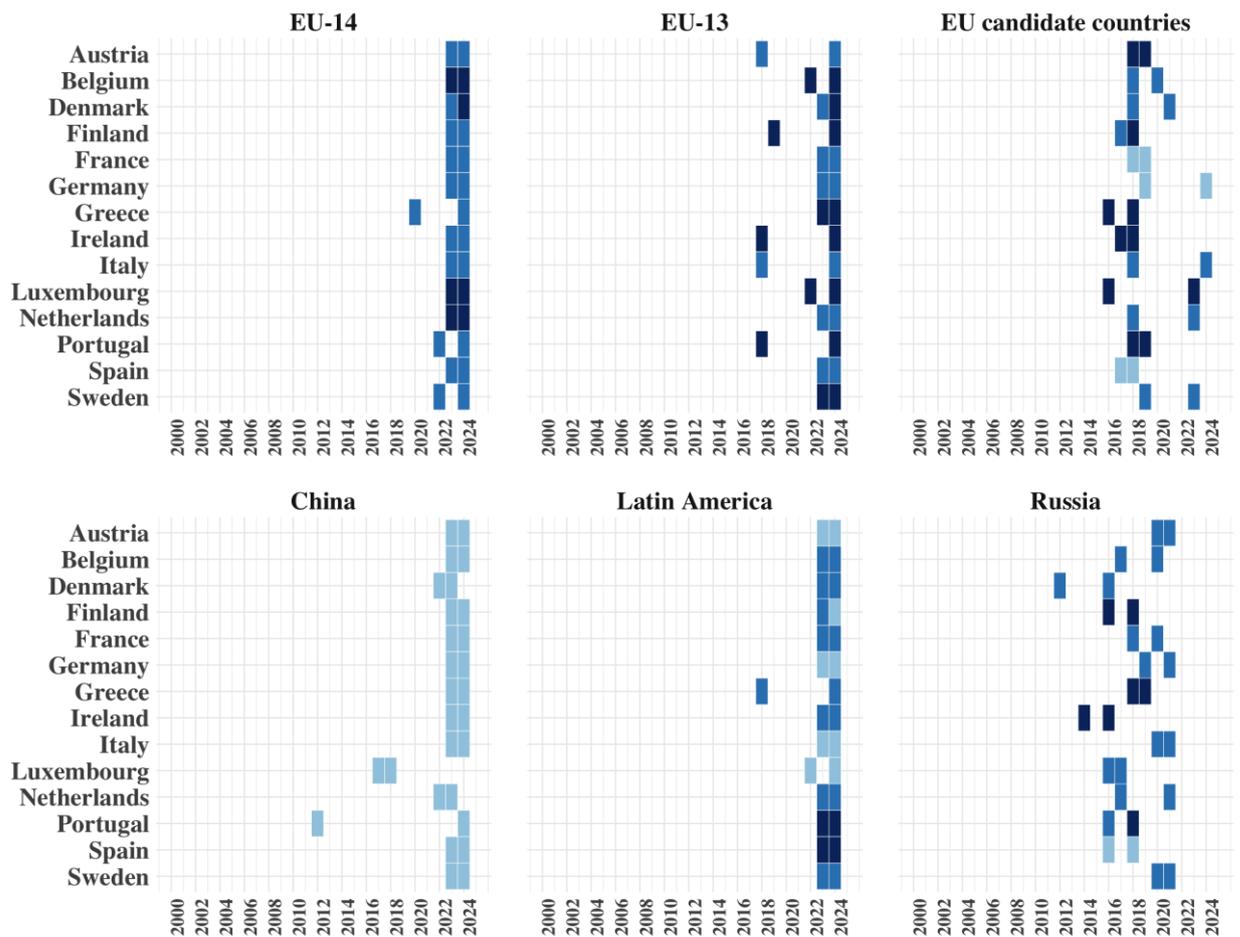

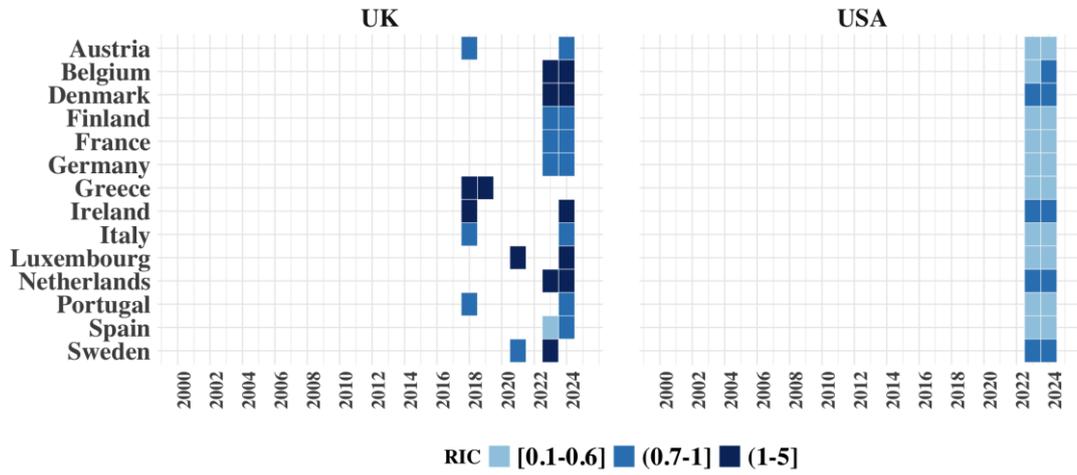

**Figure 3**. The two highest RIC peaks: EU-14

Figure 3 shows the two highest RIC peaks for EU-14 by collaboration partner. Intra-EU-14 collaboration follows a pattern of cumulative network consolidation, with most countries reaching peak RIC values at or above the expected level in 2023–2024. The Netherlands and Belgium exhibit the highest collaboration intensities, highlighting their structurally central role within the intra-EU collaboration network. A comparable gradual increase in collaboration intensity is observed between EU-14 and EU-13, with some RIC peaks occurring in 2018, reflecting the positive effect of Horizon 2020 on the ongoing deepening of intra-EU scientific integration.

    In contrast, for collaboration between EU-14 and EU candidate countries, most RIC peaks occur in 2018, with peaks for France, Germany, and Spain not exceeding 0.6. Although the 2018 peaks indicate positive effects of Horizon 2020, the subsequent decline in collaboration intensity suggests instability in collaboration between EU-14 and candidate countries and a strong dependence on EU funding.

    EU-14 collaboration with China demonstrates consistent growth, but peak RIC values do not exceed 0.43, indicating expanding but still comparatively peripheral integration relative to intra-European collaboration. In contrast, collaboration with Latin America achieves higher intensity levels, particularly for Spain and Portugal, highlighting the continued influence of historical, linguistic, and cultural proximity in shaping international scientific linkages.

    Collaboration with Russia exhibits a temporally bounded integration phase, with peak intensities concentrated between 2016 and 2021, followed by a decline consistent with the start of full-scale Russo-Ukrainian war. Most RIC peaks with Russia reach or exceed the expected level, with RIC peaks above 1.0 having Finland, Greece, and Ireland.

    Finally, collaboration with the UK and the USA shows peak intensities concentrated in 2023–2024. EU-14 countries demonstrate both increasing and relatively strong integration with the UK, while collaboration with the USA, despite temporal growth, remains comparatively moderate in intensity with RIC peaks not exceeding 0.7.

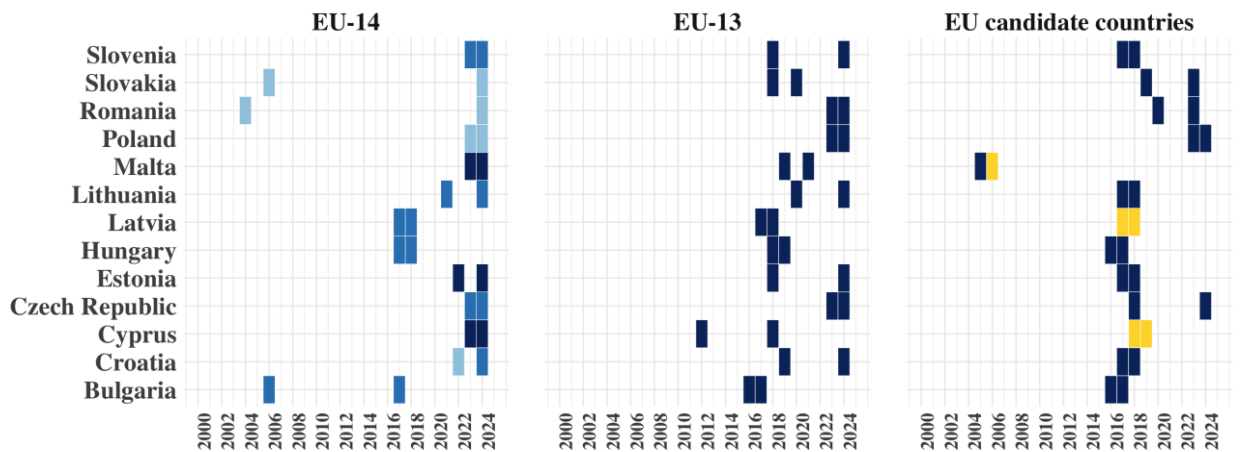

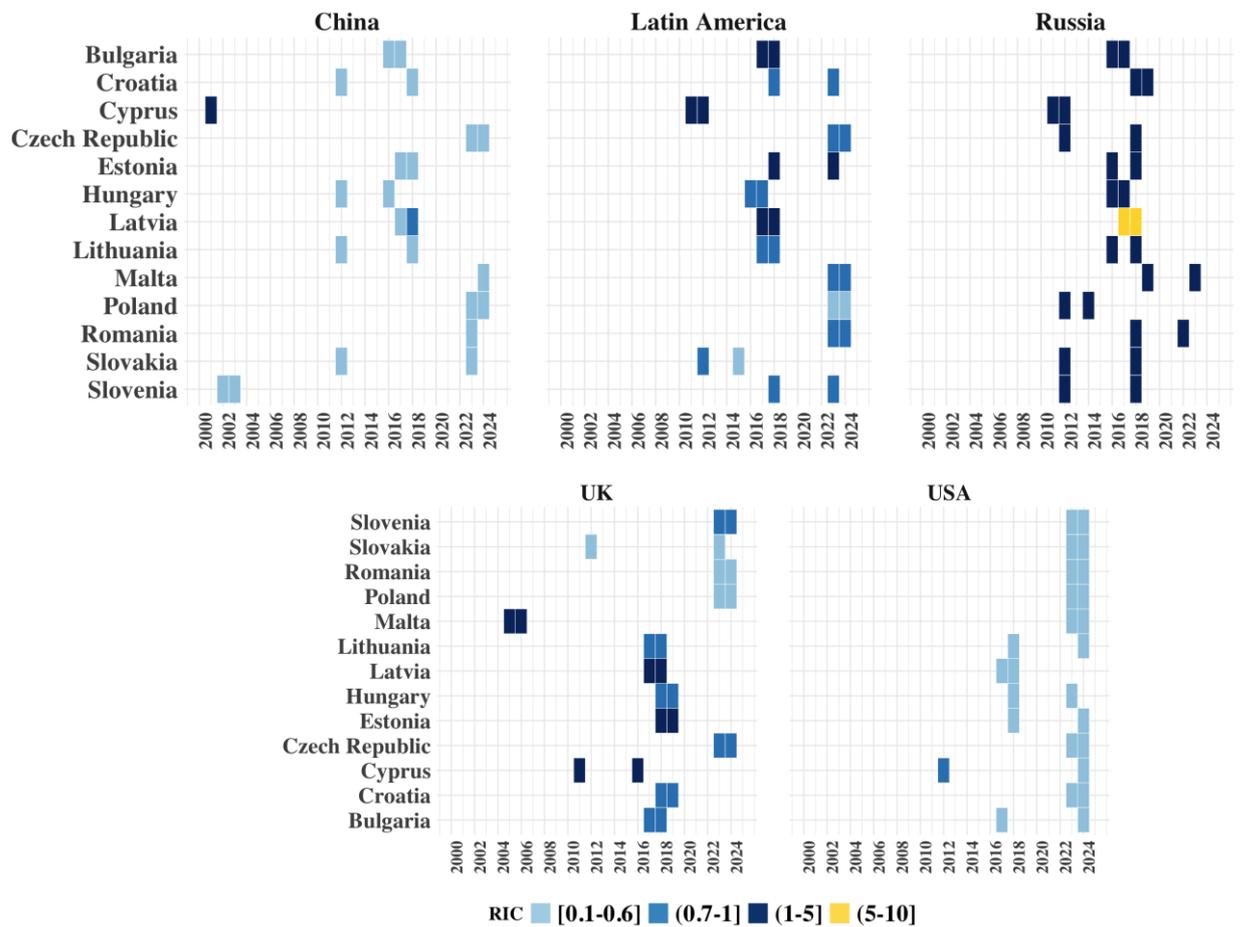

**Figure 4.** The two highest RIC peaks:EU-13

Figure 4 shows that collaboration between EU-13 countries and EU-14 broadly mirrors patterns observed for EU-14 collaboration with EU-13. RIC peaks are primarily concentrated in 2023–2024, with some occurring around 2018. The concentration of peak intensities toward the end of the observation period suggests progressive consolidation of intra-EU research network. However, the presence of countries exhibiting RIC peaks exclusively in 2018, as well as cases where peak values remain below 0.7, indicates heterogeneous integration trajectories across EU-13 countries, reflecting differences in national research capacity and network embeddedness.

Intra-EU-13 collaboration displays a more fragmented and temporally dispersed structure. Although most RIC peaks exceed the expected level, their temporal clustering around 2018 and 2023–2024 indicates episodic collaboration dynamics, suggesting limited structural consolidation within intra-EU-13 research networks.

Collaboration with EU candidate countries demonstrates comparatively higher intensity levels, with most RIC peaks concentrated around 2018. This temporal concentration suggests a stronger reliance on EU Framework Programme funding in shaping EU-13 collaboration with candidate countries relative to intra-EU collaboration dynamics.

In contrast to the gradual increase in collaboration intensity observed between EU-14 and China, EU-13 collaboration with China appears more episodic, with RIC peaks occurring around 2012, 2017–2018, and 2023–2024. These temporal clusters further suggest a strong dependence on EU Framework Programme funding facilitating collaboration between EU-13 and China. In most cases, RIC peaks remain below 0.7, indicating comparatively limited structural integration.

Collaboration with Latin America demonstrates higher intensity levels, compared to China with RIC peaks exceeding the expected level for Bulgaria, Cyprus, Estonia, and Latvia. Similar to collaboration with China, RIC peaks intensities are concentrated within the same three temporal clusters.

Collaboration with Russia exhibits a temporally bounded pattern consistent with that observed for EU-14 countries. All recorded RIC peaks exceed the expected level, with most concentrated between 2012 and 2018. Isolated RIC peaks observed for Romania in 2022 and Malta in 2023 indicate residual collaboration activity. Overall, the concentration of high RIC values prior to 2022 suggests that the Russo-Ukrainian war has affected well-established scientific collaboration channels.

Collaboration with the UK follows a pattern similar to that observed with other partners, with RIC peaks clustering around 2018 and in 2023-2024. In most cases the observed RIC peaks are rather high, except for Slovakia, Romania and Poland. Collaboration with the USA demonstrates temporal growth but RIC peaks range from 0.6 for Cyprus to 0.31 for Poland and 0.27 for Slovakia.

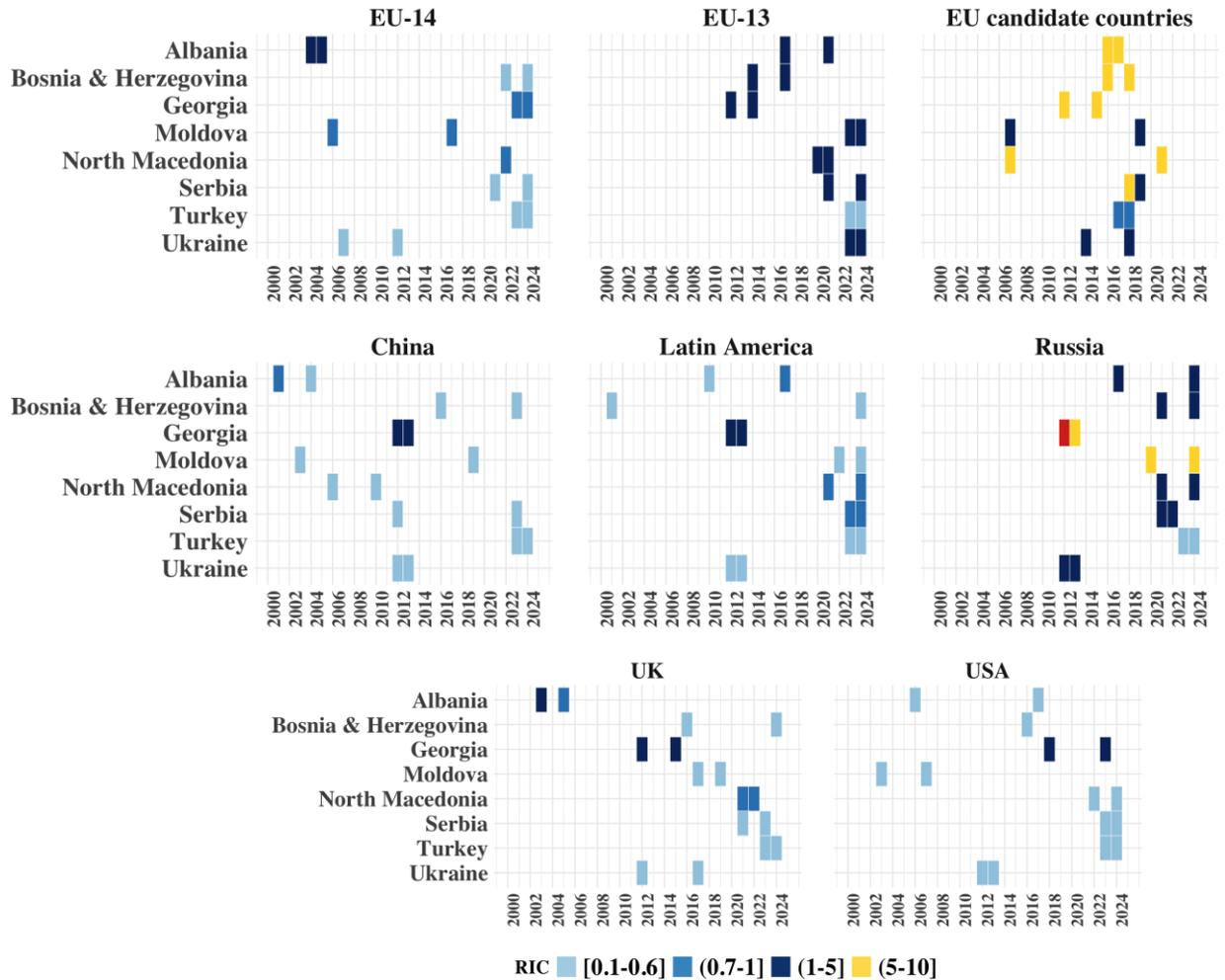

**Figure 5.** The two highest RIC peaks: EU candidate countries

Figure 5 presents the two highest RIC peaks for EU candidate countries by collaboration partner. Collaboration with EU-14 countries demonstrates temporally dispersed peak intensities between 2004 and 2024, with only a few countries reaching peaks in 2024. Georgia represents the most pronounced case of gradual increase in collaboration with EU-14, achieving relatively high RIC of 0.86 in 2024. In contrast, Ukraine has its RIC peak of 0.37 in 2012. Moldova has its recent RIC peak of 0.83 in 2017, suggesting earlier phases of integration that have not been sustained at comparable levels.

Collaboration with EU-13 countries generally exhibits higher RIC peak values across most EC candidate countries, indicating stronger relational proximity and potentially lower structural barriers to collaboration. However, sustained growth patterns are observable only for Moldova, Serbia, Turkey, and Ukraine, with RIC peaks concentrated in 2023–2024.

Collaboration among EU candidate countries demonstrates even higher RIC peaks compared to collaboration with EU-13 partners. However, these peaks are strongly concentrated around 2018, indicating an episodic high intensity rather than a well-established collaboration. This temporal clustering suggests that collaboration among candidate countries may be particularly sensitive to project-based funding mechanisms and regional policy initiatives.

Consistent with patterns observed for EU-14 and EU-13, collaboration between EU candidate countries and China remains limited in intensity. With the exception of Georgia, which exceeds the expected collaboration level in 2012, most RIC peak values remain comparatively low.

Collaboration with Latin America shows moderate but gradually increasing intensity for several EU candidate countries, with RIC peak values slightly exceeding those observed for China. Collaboration with the UK demonstrates moderate integration, with peak intensities concentrated in the late 2010s and early 2020s. Collaboration with the USA follows a similar temporal expansion and also remains comparatively moderate in intensity. These patterns indicate expanding diversification of external collaboration portfolios, although such engagement remains less institutionalised than collaboration within the group and with EU-13 countries.

In collaboration with Russia, Ukraine and Georgia exhibit RIC peaks only in 2012, whereas for other countries RIC peaks are distributed across the 2020–2024 period. RIC peaks exceed the expected level for all countries except Turkey. Particularly high values are observed for Moldova in 2024.

Notably, Ukraine exhibits RIC peaks with most collaboration partners in 2012, with additional RIC peak observed for collaboration with EU-13 countries in 2023–2024 and with EU candidate countries in 2018. The 2012 RIC peaks may be associated with the final years of FP7, which likely intensified publication output. The gradual increase in collaboration with EU-13 countries suggests growing research integration with the EU, primarily mediated through partnerships with these countries rather than with EU-14.

Similarly, Moldova has intensified collaboration with EU-13 countries rather than with EU-14 in recent years, while also maintaining strong collaboration with Russia. Georgia also exhibits RIC peaks with most partners in 2012; however, in contrast to Ukraine, its collaboration with EU-14 countries and the United States shows peak intensities in more recent years.

## 5 Discussion and conclusions

This study examined the dynamics of scientific collaboration among EU-14, EU-13, EU candidate countries, as well as their collaboration with selected global partners by analysing the Relative Intensity of Collaboration (RIC) from 2000 to 2024.

The study findings align with two established streams of research on scientific collaboration. First, collaboration patterns are influenced by historical ties, socioeconomic, geographical, cultural, and linguistic proximity (Hoekman et al., 2010; Scherngell & Barber, 2009; Hoekman et al., 2010; Frenken et al., 2009; Glänzel & Schubert, 2004; European Commission, 2025; Zitt et al., 2000). Second, collaboration is shaped by targeted funding instruments (Álvarez-Bornstein & Bordons, 2021; Barre et al., 2013), including both national and supranational mechanisms, such as the EU Framework Programmes. These theoretical perspectives provide a lens for interpreting the observed patterns of integration, variability, and resilience in European and global collaboration networks.

Overall, the analysis indicates increasing scientific integration both within the European groups and between them and selected global partners. However, there are persistent regional clustering and differentiated collaboration structures within the European research system. EU-14 countries form the core of the collaboration network, characterised by strong intra-group ties and intensive collaboration with EU-13 countries, EU candidate countries, and the UK. EU-13 countries occupy an intermediate position, maintaining moderate collaboration with EU-14 countries while demonstrating substantially stronger collaboration within their own group and with EU candidate countries and Russia. Their collaboration with global partners, including the UK and the USA, is less intensive than that of EU-14 countries but remains stronger than that observed for EU candidate countries. EU candidate countries display comparatively weaker collaboration with EU-14 countries and global partners. Instead, their collaboration patterns are primarily oriented toward intra-group cooperation and partnerships with EU-13 countries and Russia, reflecting lower levels of integration into the core European and transatlantic collaboration structures. This aligns with findings obtained by Marini (2023) on comparison of EU-13 and EU candidate countries.

EU-14 countries exhibit the most stable and cohesive collaboration patterns across all collaboration partners. In contrast, collaboration intensity among EU-13 countries shows greater fluctuation, while even stronger variability is observed for EU candidate countries. The RIC peaks in collaboration intensity observed for EU-13 and EU candidate countries in 2012 and 2018 coincide with the final phases of EU Framework Programmes, suggesting the influence of Horizon 2020 and Horizon Europe funding cycles. Notably, RIC peaks in these years are also observed in collaboration with global partners and in collaboration between EU-14 and both EU-13 and EU candidate countries.

On the one hand, these findings highlight the positive role of EU Framework Programmes in strengthening international scientific collaboration, which is consistent with previous research (Makkonen & Mitze, 2016; Mattsson et al., 2008; Kwiek, 2021; Teodorescu & Andrei, 2011; European Commission, 2021; Barre et al., 2013). On the other hand, the results reveal the relative fragility of international collaboration involving EU-13 and EU candidate countries, indicating their stronger dependence on EU funding mechanisms and potentially weaker national research funding capacity, with this issue being more urgent for EU candidate countries.

The growing collaboration between EU-14 and EU-13 countries with the UK despite Brexit may also be partly explained by the UK's continued association with Horizon 2020 and Horizon Europe, as well as by long-standing collaborative research networks. However, previous research (Oldac & Olivos, 2025) suggests a shift in the UK's collaboration patterns, marked by reduced EU engagement and growing research and funding ties with China. Growing collaboration is also observed with Latin America, although trends in collaboration intensity differ between the groups. For EU-14 countries, collaboration with Latin America demonstrates gradual and sustained growth, with particularly strong links observed for Spain and Portugal, likely reflecting cultural and linguistic proximity. In contrast, EU-13 and EU candidate countries display more volatile collaboration trajectories with Latin American partners. In turn, the growing but still low collaboration with China, with similar intensity across all three groups can be attributed to the absence of long-standing research networks and limited cultural and linguistic proximity.

The influence of long-standing collaboration ties is also reflected in the resilience of scientific collaboration networks involving Russia. This is evidenced by an increase in collaboration following 2014 and only a marginal decline after 2022, with collaboration intensity exceeding that observed between EU-14 with the USA. This is consistent with findings reported by Zhang et al. (2024). Although Russian entities were suspended from Horizon Europe in 2022 and ongoing grants were terminated, collaboration sustained through personal research networks may have persisted.

This study has several limitations. First, the analysis relies exclusively on CWTS Scopus data. Second, collaboration intensity was calculated using the full counting method, which reflects participation in collaboration rather than proportional contribution; fractional counting could provide alternative insights (Cao et al., 2025).

To summarise, sustaining and deepening international collaboration requires consistent policy attention, targeted funding mechanisms, and long-term strategic commitment from both the European Union and national governments.

**Declaration**
The author claims no conflict of interest